\documentclass[aps]{revtex4}
\begin{document}
\title{Sewing sound quantum flesh onto classical bones}
\author{Edward D. Davis}
\email{davis@kuc01.kuniv.edu.kw}
\affiliation{Department of Physics, Kuwait University, P.O. Box 5969, Safat, Kuwait}
\date{\today}
\begin{abstract}
Semiclassical transformation theory implies an integral representation for stationary-state wave functions $\psi_m(q)$ in 
terms of angle-action variables ($\theta,J$). It is a particular solution of Schr\"odinger's time-independent equation 
when terms of order $\hbar^2$ and higher are omitted, but the pre-exponential factor $A(q,\theta)$ in the integrand of
this integral representation does not possess the correct dependence on $q$. The origin of the problem is identified: the 
standard unitarity condition invoked in semiclassical transformation theory does not fix adequately in $A(q,\theta)$ a 
factor which is a function of the action $J$ written in terms of $q$ and $\theta$. A prescription for an improved 
choice of this factor, based on succesfully reproducing the leading behaviour of wave functions in the vicinity of potential
minima, is outlined. Exact evaluation of the modified integral representation via the Residue Theorem is possible. It 
yields wave functions which are not, in general, orthogonal. However, closed-form results obtained after Gram-Schmidt 
orthogonalization bear a striking resemblance to the exact analytical expressions for the stationary-state wave functions 
of the various potential models considered (namely, a P\"{o}schl-Teller oscillator and the Morse oscillator).
\end{abstract}
\maketitle


Semiclassical transformation theory approximates various quantum probability amplitudes in terms of the generating 
functions of classical canonical transformations~\cite{Mi74}. The theory is aesthetically very satisfying in 
the manner in which it exploits parallels between quantum mechanics and classical mechanics. It has also proved eminently 
useful. The basic relations for amplitudes in terms of generating functions, although more than 30 years old, are still 
central to current research~\cite{Mi01,Ka01}. 

A formal application of semiclassical transformation theory~\cite{Mi74} implies that the bound states of a conservative
system of one degree of freedom have actions $J_m$ ($m=0,1,2,\ldots$) given by the appropriate Bohr-Sommerfeld quantization 
condition and approximate configuration space wave functions
\begin{equation}\label{Mwf}
\psi_m(q)=N_m\int A(q,\theta)\exp\left[(i/\hbar)F_{cl}(q,\theta)\right]\phi_m(\theta)d\theta,
\end{equation}
where $F_{cl}(q,\theta)$ is the generating function of the first kind (see chapter 9 in \cite{Go80}) for the canonical 
transformation from conjugate cartesian variables ($q,p$) to the angle-action variables ($\theta,J$) for the system, the 
pre-exponential factor
\begin{equation}\label{MA}
A(q,\theta)={\cal A}_{U}(q,\theta)\equiv\left[\frac{-1}{2\pi i\hbar}\frac{\partial^2 F_{cl}}{\partial q\partial\theta}
\right]^{1/2}, 
\end{equation}
$\phi_m(\theta)\equiv(2\pi i\hbar)^{-1/2}\exp(iJ_m\theta/\hbar)$ and $N_m$ is a normalization constant~\cite{Phase}.
Equations (\ref{Mwf}) and (\ref{MA}) reduce to an acceptable result (namely, the WKB approximation) if the integral is 
evaluated in the $\hbar\rightarrow 0$ stationary phase approximation. However, if one attempts to go beyond the stationary
phase approximation as contemporary studies do, then there are problems. As specialization of 
Eqs.~(\ref{Mwf}) and (\ref{MA}) to various examples reveals (see below and Appendix C.2 in~\cite{Ch91}), the $q$ 
dependence of ${\cal A}_U$ is typically spurious. In particular, wave functions do not have well-defined parity under 
circumstances when this symmetry is expected.

The integral representation of Eq.~(\ref{Mwf}) with the pre-exponential factor in Eq.~(\ref{MA}) is a {\em particular\/} 
solution of Schr\"odinger's time-independent equation when terms of order equal to or greater than $\hbar^2$ are discarded, 
but the {\em general\/} solution (to this order) only constrains the pre-exponential factor $A(q,\theta)$ to be of the 
form~\cite{DG03}
\begin{equation}\label{GA} 
{\cal A}_{G}(q,\theta)\equiv f\left(J(q,\theta)\right){\cal A}_{U}(q,\theta),
\end{equation}
where $f$ is an arbitary differentiable function of $J(q,\theta)=-\partial F_{cl}\left/\partial\theta\right.
$~\cite{GenSol}. I claim that, in the case of potentials displaying a single minimum, the factor $f\left(J(q,\theta)\right
)$ can be chosen so that the corresponding approximate wave functions have the desired $q$ behaviour in the vicinity of the
potential's minimum. In the case of even potentials $V(q)$ (with a single minimum), this is enough to guarantee 
that wave functions have the right parity. No matter what choice of $f$ is made, the pre-exponential factors ${\cal A}_{U}
(q,\theta)$ and ${\cal A}_{G}(q,\theta)$ are indistinguishable if the $\hbar\rightarrow 0$ stationary phase approximation 
is invoked [because $J(q,\theta)$ is then replaced by the constant $J_m$].

The choice of pre-exponential factor ${\cal A}_U(q,\theta)$ guarantees that, in the limit $\hbar\rightarrow 0$, the 
corresponding kernel ${\cal K}_U(q,\theta)\equiv{\cal A}_U(q,\theta)\exp[(i/\hbar)F_{cl}(q,\theta)]$ in Eq.\ (\ref{Mwf}) 
is an element of a unitary transformation matrix (or a quantum probability amplitude). 
Although this is an appealing property, it is by no means obvious that it is appropriate as 
there is no rigorous unitary quantum mechanical counterpart to transformations from cartesian phase space variables to 
(conventional) angle-action variables. In fact, one implication of the previous paragraph is that it is permissable to 
insist that ${\cal K}(q,\theta)\equiv A(q,\theta)\exp[(i/\hbar)F_{cl}(q,\theta)]$ is an element of a unitary transformation
matrix in the limit $\hbar\rightarrow 0$. The upshot of the considerations below is that this unitarity condition must be 
supplemented by more information to fix the pre-exponential factor completely.


Support for the assertion above about the role of the factor $f\left(J(q,\theta)\right)$ comes from consideration of some 
analytically soluble models: the simple harmonic oscillator (discussed in~\cite{DG03}), a P\"{o}schl-Teller oscillator 
described by the potential $V(q)=V_0\tan^2(\pi q/a)$ ($V_0>0$) and the Morse oscillator with potential $V(q)=D[1-\exp(-q/d)
]^2$ ($D,d>0$). The P\"{o}schl-Teller and Morse models, which are taken up in the present work, also force one to consider 
issues related to the contour of integration in Eq.~(\ref{Mwf}).

For the example of the harmonic oscillator, eigenfunctions can be obtained by integrating over the full range of 
$\theta$ (modulo $2\pi$) for which $F_{cl}(q,\theta)$ exists for the value of $q$ under consideration~\cite{thetaHO}. 
Despite the apparent reasonableness of this prescription, it is inappropriate in the case of the P\"oschl-Teller potential.
The corresponding limits of integration would contain a $q$-dependence~\cite{thetaPT} incompatible with the requirement
that the resulting integral representation for eigenfunctions satisfies Schr\"odinger's time-independent equation when 
terms of order $\hbar^2$ and higher are dropped. (The $q$-dependent limits would give rise to error terms of order $\hbar
$.)

What alternatives are there? The general formal considerations of~\cite{DG03} suggest that the kernel ${\cal K}(q,\theta)
= A(q,\theta)\exp[(i/\hbar)F_{cl}(q,\theta)]$ in Eq.~(\ref{Mwf}) can be thought as a generating function for 
(approximate) wave functions in the sense that they are determined (up to multiplicative constants) by the coefficients 
of an expansion of ${\cal K}(q,\theta)$ in powers of $w\equiv e^{-i\theta}$. More precisely, if the Bohr-Sommerfeld 
quantization condition $J_m=(m+\mu/4)\hbar$ applies ($m$ is a non-negative integer and $\mu$ is the Maslov index), then, 
according to Eq.~(10) in~\cite{DG03}, the expansion reads
\begin{equation}\label{Kfexp}
{\cal K}\left(q,\theta(w)\right)= w^{\mu/4}\sum_{m=0}^\infty \frac{\psi_m(q)}{\widetilde{n}_m} w^m,
\end{equation}
where $\theta(w)=i\ln w$ [with $|w|$ chosen so that the series in Eq.~(\ref{Kfexp}) converges] and the $\widetilde{n}_m$'s
are constants [introduced in the denominator to simplify Eqs.~(\ref{Schint}) and (\ref{Rod})]. The expansion in 
Eq.~(\ref{Kfexp}) implies that approximate wave functions can be written as the Schl\"{a}fli-like contour integral
\begin{equation}\label{Schint}
\psi_m(q)=\frac{\widetilde{n}_m}{2\pi i}\oint\left(\frac{1}{w}\right)^{m+1+\mu/4}A\left(q,\theta(w)\right)\exp\left[
(i/\hbar)F_{cl}\left(q,\theta(w)\right)\right]dw,
\end{equation}
where the (counter-clockwise) contour of integration is a simple closed curve around a pole of order $m+1$ at $w=0$ which 
does not enclose any other singularities of the integrand.

Despite appearances, the integral representations in Eqs.~(\ref{Mwf}) and (\ref{Schint}) are closely related: since
$J_m=(m+\mu/4)\hbar$ in $\phi_m(\theta)$, the integrand of Eq.~(\ref{Mwf}) transforms into the integrand of 
Eq.~(\ref{Schint}) under the replacement of $\theta$ by $w=e^{-i\theta}$ as the variable of integration. Thus, implicit in 
Eq.~(\ref{Schint}) is a choice of the contour of integration for the integral representation in Eq.~(\ref{Mwf}). If 
possible, it should pass through the point(s) of stationary phase which yield the WKB approximation. I shall not attempt 
to be more explicit here, because, for the purposes of this paper, the Rodrigues' formula implied by Eq.~(\ref{Kfexp})
[or the Residue Theorem applied to Eq.~(\ref{Schint})], namely  
\begin{equation}\label{Rod}
\psi_m(q)=\frac{\widetilde{n}_m}{m!}\left.\frac{\partial^m\ }{\partial w^m}\left[w^{-\mu/4}{\cal K}\left(q,\theta(w)\right)
\right]\right|_{w=0},
\end{equation}
is more useful than either of the above integral representations. 

When one can distinguish between states of different parity (as in the P\"oschl-Teller model), it is necessary to introduce
separate kernels ${\cal K}_\varrho$ for the positive ($\varrho=0$) and negative parity ($\varrho=1$) states 
[obtained by making different choices of the factor $f\left(J(q,\theta)\right)$]. An expansion in powers of $z\equiv 
e^{-i2\theta}$ is now appropriate, the expansion being 
\begin{equation}\label{Kepinz}
{\cal K}_\varrho\left(q,\theta(z)\right)= z^{(\mu/4+\varrho)/2}\sum_{m=0}^\infty \frac{\psi_{2m+\varrho}(q)}{\widetilde{
n}_{2m+\varrho}} z^m.
\end{equation}
The associated Rodrigues' formula expresses $\psi_{2m+\varrho}(q)$ in terms of the $m$th [{\em not\/} the 
$(2m+\varrho)$th] partial derivative of $z^{-(\mu/4+\varrho)/2}{\cal K}_\varrho\left(q,\theta(z)\right)$ with respect to 
$z$ (evaluated at $z=0$). A division into two classes of states can also be helpful for models which do not respect parity 
(see the discussion of the Morse oscillator below).

The harmonic oscillator problem furnishes evidence that expansions like that in Eq.~(\ref{Kfexp}) or Eq.~(\ref{Kepinz}) do,
in fact, exist. Consistent with Eq.~(20) in \cite{DG03}, the kernels of definite parity are
\begin{equation}
{\cal K}_\varrho(q,\theta) = y^\varrho (\cos\theta)^{-(1/2+\varrho)}\exp\left(-\frac{i}{2}y^2\tan\theta\right),
\end{equation}
where the dimensionless variable $y\equiv\sqrt{m\omega/\hbar}\,q$ ($m$ is the mass of the oscillator and $\omega$ its 
angular frequency). The expression which results on the substitution of $e^{-i2\theta}$ by $z$ can be recast into the form
\begin{equation}\label{HOKw}
{\cal K}_\varrho = 2^{1/2+\varrho}z^{1/4+\varrho/2}e^{-y^2/2}\exp\left(-\frac{1}{4}\frac{\partial^2\ }{\partial y^2}\right)
y^\varrho e^{zy^2}
\end{equation}
suitable for expansion in powers of $z$~\cite{FTtrick}. Using the Maclaurin series in $z$ for $y^\varrho e^{y^2z}$ and the 
fact that the Hermite polynomial $H_n(y)=2^n\exp(-\frac{1}{4}\frac{\partial^2\ }{\partial y^2})y^n$ (see Eq.~(1) 
in \cite{Fe98}), Eq.~(\ref{HOKw}) implies that
\begin{equation}\label{HOKfin}
{\cal K}_\varrho = 2^{(\varrho+1)/2}\left(\frac{\pi\hbar}{m\omega}\right)^{1/4}z^{1/4+\varrho/2}\sum_{m=0}^\infty \frac{[
(2m+\varrho)!]^{1/2}}{2^m m!}\,\varphi_{2m+\varrho}(q) z^k,
\end{equation}
where the $\varphi_n(q)$'s are the exact normalized energy eigenfunctions of the simple harmonic oscillator ($\varphi_n=
[m\omega/(\pi\hbar)]^{1/4}(2^n n!)^{1/2}H_n(y)e^{-y^2/2}$). Since the Maslov index $\mu=2$ for the harmonic oscillator, 
Eq.~(\ref{HOKfin}) constitutes an explicit realization of Eq.~(\ref{Kepinz}).


The P\"{o}schl-Teller oscillator is instructive, because unlike the example of the harmonic oscillator, the results 
obtained are not exact and so it is possible to assess the quality of the approximation based on Eqs.~(\ref{Mwf}) and 
(\ref{GA}). Although the relevant transformation from cartesian phase variables to angle-action variables appears in 
textbooks (see, for example, chapter 7 in \cite{PR82}), the corresponding generating function of the first kind is not so 
readily available. The calculation of this generating function as the Legendre transform of Hamilton's characteristic 
function $W(q,J)$ for the P\"{o}schl-Teller problem is straightforward but lengthy, it being necessary to treat separately 
each branch of the multiple-valued $W(q,J)$ for both signs of $q$. Fortunately, the result can be compactly expressed [in 
a form suitable for use in the Rodrigues' formula implied by Eq.~(\ref{Kepinz})] as follows:
\begin{equation}\label{FPT}
F_{cl}(q,\theta) = i J_s\ln\left\{\left.\left[\sqrt{(1-e^{-i2\theta})^2+4\cos^2(\pi q/a)e^{-i2\theta}}+(1-e^{-2i\theta})
\right]\right/\left[2\cos(\pi q/a)\right]\right\}
\end{equation}
where the parameter $J_s\equiv a\sqrt{2mV_0}/\pi$ sets the scale for classical actions ($m$ is again the mass of the
oscillator). In Eq.~(\ref{FPT}), which holds for both signs of $q$ and the pertinent ranges of $\theta$ 
(see~\cite{thetaPT}), it is understood that the 
positive branch of the square root and the principal value of the logarithm function are to be adopted~\cite{FPTalt}. 
Partial differentiation of $F_{cl}(q,\theta)$ with respect to $\theta$~\cite{ThetaPD} yields
\begin{equation}\label{Jqtheta}
J(q,\theta) \equiv -\frac{\partial F_{cl}}{\partial\theta} = J_s\left\{\left(1-\left[\left.\sin(\pi q/a)\right/\cos\theta
\right]^2\right)^{-1/2}-1\right\},
\end{equation} 
implying that, for the present problem, the factor $f(J(q,\theta))$ in ${\cal A}_G$ is, in effect, a function of the 
combination $\sin(\pi q/a)/\cos\theta$. Likewise, the second partial derivative of $F_{cl}$ required in 
${\cal A}_G$ (and ${\cal A}_U$)
\begin{equation}\label{SDF}
\frac{\partial^2 F_{cl}}{\partial q\partial\theta} = -\frac{\pi}{a}J_s \cos(\pi q/a)\frac{\sin(\pi q/a)}{\cos^2\theta}
\left\{1-\left[\left.\sin(\pi q/a)\right/\cos\theta\right]^2\right\}^{-3/2} 
\end{equation}
amounts to the product of $\cot(\pi q/a)$ with a function of $\sin(\pi q/a)/\cos\theta$. Pre-exponential factors must, 
therefore, be of 
the form $[\cot(\pi q/a)]^{1/2}G(\sin(\pi q/a)/\cos\theta)$, where $G$ is some differentiable function. This functional 
form is compatible with choices which possess the leading small $q$ behaviour expected of energy eigenfunctions of 
definite parity, the {\em simplest\/} being
\begin{equation}\label{SimCh}
A_\varrho(q,\theta)=[\cot(\pi q/a)]^{1/2}\left(\frac{\sin(\pi q/a)}{\cos\theta}\right)^{1/2+\varrho}
                   =[\cos(\pi q/a)]^{1/2}\,\frac{\sin^\varrho(\pi q/a)}{(\cos\theta)^{1/2+\varrho}},
\end{equation}
where, as above, $\varrho=0$ (1) for positive (negative) parity states. Equation~(\ref{SDF}) serves to illustrate that 
the pre-exponential factor ${\cal A}_U$ would have the wrong leading behaviour for small $q$.

With the pre-exponential factors of Eq.~(\ref{SimCh}), the Rodrigues' formula for approximate energy eigenfunctions implies
that
\begin{equation}\label{nonOrt}
\psi_{2m+\varrho}(q)\propto\left[\cos\left(\frac{\pi}{a}q\right)\right]^{\Lambda+1/2}P_m^{(\varrho)}\left(\sin(\pi q/a)
\right),
\end{equation}
where $\Lambda\equiv J_s/\hbar$ and
\begin{equation}
P_k^{(\varrho)}(x)\equiv x^\varrho\left.\frac{d^k\ }{dz^k}\left[(1+z)^{-(1/2+\varrho)}\left(1-z+\sqrt{(1+z)^2-4x^2z}
\right)^{-\Lambda}\right]\right|_{z=0}
\end{equation}
is a polynomial of order $2k+\varrho$ and well-defined parity [$P_k^{(\varrho)}(-x)=(-1)^\varrho P_k^{(\varrho)}(x)$]. 
Calculation shows that these approximate wave functions are {\em not\/} orthogonal although the degree of 
non-orthogonality is slight if $\Lambda\gg 1$~\cite{Overlap}. In terms of $x=\sin\left(\pi q/a\right)$, orthogonalization 
amounts to the construction of a set of orthogonal polynomials on the interval $-1<x<+1$ with weight function 
$(1-x^2)^\Lambda$. This is the set of ultraspherical polynomials $C^{(\Lambda+1/2)}_n(x)$~\cite{AS64}. Accordingly, after 
orthonormalization, the approximate wave functions are
\begin{equation}\label{orthappwvf}
N_n(\Lambda)\left[\cos(\pi q/a)\right]^{\Lambda+1/2}C_n^{(\Lambda+1/2)}\left(\sin(\pi q/a)\right),
\end{equation}
where the $N_n(\Lambda)$'s are normalization constants ($n=0,1,2,\ldots$). Remarkably, the exact energy eigenfunctions are 
also given by Eq.~(\ref{orthappwvf}) with $\Lambda$ replaced by $\sqrt{\Lambda^2+1/4}$.

The analytic expressions for the non-orthogonal wave functions in Eq.~(\ref{nonOrt}) and the orthogonalized wave functions 
in Eq.~(\ref{orthappwvf}) do not appear to be significantly different: barring the overall normalization factors, the 
coefficients of each power of $x=\sin(\pi q/a)$ agree to leading order in $\Lambda$ (see~\cite{Mo02} for some examples). 
Nevertheless, orthogonalization is crucial. Numerical studies reveal that it improves substantially the agreement between 
approximate and exact eigenfunctions (so that it is good provided $\Lambda^2\gg 1$). Perhaps more telling is the fact that,  
even when $\Lambda\gg 1$, the non-orthogonal wave functions can be inferior to (approximately) normalized WKB 
wavefunctions~\cite{WKBnrm} inside the classically-allowed region (well away from the classical turning points).


Analysis of the Morse oscillator serves to illustrate that it is possible to treat models which do not possess wave 
functions of good parity. The guiding principle I have found useful is to insist that, in the immediate vicinity of the 
potential minimum (at $q=0$), the approximate Morse oscillator wave functions $\widetilde{\psi}_n(q)$ resemble those of a 
harmonic oscillator of the same mass $m$ and the appropriate angular frequency, namely $\widetilde{\omega}=\widetilde{J}_s/
(m{d}^{2})$ where the action $\widetilde{J}_s\equiv\sqrt{2mD}d$~\cite{ExactHO}. Not only is the pre-exponential factor 
$A(q,\theta)$ at issue but also the analytic continuation to complex values of $\theta$ of the generating function 
$F_{cl}(q,\theta)$ to be used in Eqs.~(\ref{Schint}) or (\ref{Rod}) and even the choice of $F_{cl}(q,\theta)$ for real 
values of $\theta$~\cite{MValue}. Adoption of the $q>0$ result for $F_{cl}(q,\theta)$ is indicated, which, for $0<\theta<
\pi$, reads
\begin{equation}\label{FMorse}
F_{cl}(q,\theta)=\widetilde{J}_s\left\{\sqrt{\lambda^2(q,\theta)-(1-e^{-q/d})^2}+\mbox{Arccos}\left[(1-e^{-q/d})/
\lambda(q,\theta)\right]-\theta\right\},
\end{equation}
where $\lambda(q,\theta)$ is the positive root of the relation $q/d=\ln[(1+\lambda\cos\theta)/(1-\lambda^2)]$ and Arccos 
denotes the principal value of the inverse cosine function. It is understood that the principal values of all square 
roots present either explicitly or implicitly [in $\lambda(\theta,q)$] are to be used.
The analytic continuation (of interest) of $F_{cl}(q,\theta)$ to complex $\theta$ is such that the alternative branch of the 
square root $\sqrt{\lambda^2(q,\theta)-(1-e^{-q/d})^2}$ in Eq.~(\ref{FMorse}) is adopted~\cite{ConHO}. 

In view of the fact that I compare Morse oscillator wave functions with those of the harmonic oscillator, it is still 
useful to divide them into a $\varrho=0$ class $\{\widetilde{\psi}_{2m}(q)\}$ and a $\varrho=1$ class $\{\widetilde{\psi}_{
2m+1}(q)\}$, which correspond (near $q=0$) to the positive and negative parity harmonic oscillator wave functions, 
respectively ($m=0,1,\ldots$). Suitable pre-exponential factors for the two classes of wave functions are $[e^{-2q}\cos^2
\theta+4(1-e^{-q})]^{-1/4}\widetilde{G}_\varrho(\lambda(q,\theta))$, where the simplest appropriate choice of 
$\widetilde{G}_\varrho(x)$ is $\widetilde{G}_\varrho(x)=x^\varrho$. For this choice, the approximate energy eigenfunctions
are found to be of the form
\begin{equation}\label{Mwave}
\widetilde{\psi}_n(q) \propto (e^{-q/d})^{\widetilde{\Lambda}-n-1/2} \exp(-\widetilde{\Lambda}e^{-q/d})
                               \widetilde{P}_n(e^{-q/d})  
\end{equation} 
where $\widetilde{P}_n(x)$ is a polynomial of order $n$ and $\widetilde{\Lambda}\equiv\widetilde{J_s}/\hbar$. As in the case
of the P\"oschl-Teller oscillator, the wave functions $\widetilde{\psi}_n(q)$ in Eq.~(\ref{Mwave}) are weakly 
non-orthogonal. After orthonormalization, the exact bound state wave functions are obtained.

In summary, the examples in this paper illustrate just how much can be learnt about wave functions of simple systems using 
a modification of the approximate integral representation in Eq.~(\ref{Mwf}). One change seeks to correct for obvious 
deficiencies of the pre-exponential factor implied by semiclassical transformation theory [namely, ${\cal A}_U$ in 
Eq.~(\ref{MA})]. Other changes may appear less welcome but are equally essential: the use of a complex-valued contour of 
integration and, sometimes, an appropriate analytic continuation of the generating function. Despite the lack of intuitive 
appeal inherent in the use of a complex-valued contour, it does facilitate analytic evaluation of the integral 
representation (since the Residue Theorem may be invoked). Typically, semiclassical integral representations improving on 
Eq.~(\ref{Mwf}) entail integration over classical trajectories~\cite{Li86,ZK96,MK98}. This is one obvious difference between
these representations and the one explored here. There is another. Corrections to the pre-exponential factor have been 
guided in this work by the structure of the general solution in Eq.~(\ref{GA}). This approach is inappropriate for integral
representations involving integration over classical trajectories: the factor $f(J(q,\theta))$ in Eq.~(\ref{GA}) becomes an 
invisible constant [since $J(q,\theta)$ is equal to the fixed action on a trajectory]. 

\begin{acknowledgments}
I would like to thank Ghassan I. Ghandour for bringing~\cite{Fe98} to my attention and for suggesting investigation of the 
P\"{o}schl-Teller model. 
\end{acknowledgments}


\begin{thebibliography}{}
\bibitem{Mi74} W. H. Miller, Adv.\ Chem.\ Phys.\ {\bf 25} (1974), 69.
\bibitem{Mi01} W. H. Miller, J.\ Phys.\ Chem.\ A 105 (2001), 2942.
\bibitem{Ka01} K. G. Kay, J.\ Phys.\ Chem.\ A 105 (2001), 2535.
\bibitem{Go80} H. Goldstein, {\sl Classical Mechanics}, 2nd.\ Ed.\ (Addison-Wesley, Reading, 1980).
\bibitem{Phase} I have adopted the phase conventions of~\cite{Mi74} in defining $A(q,\theta)$ and $\phi_m(\theta)$.
\bibitem{Ch91} M. S. Child, {\sl Semiclassical Mechanics with Molecular Applications}, (Clarendon Press, Oxford, 1991).
\bibitem{DG03} E. D. Davis and G. I. Ghandour, Phys.\ Lett.\ A 309 (2003), 1.
\bibitem{GenSol} The eigenenergies are $E_m\equiv K(J_m)$, where $K(J)$ is the classical Hamiltonian function for 
                 angle-action variables. The factor of $2\pi i\hbar$ has been included in Eq.~(\ref{GA}) to facilitate 
                 comparison with Eq.~(\ref{MA}).
\bibitem{thetaHO} This is illustrated by Eqs.~(C.62a,b) in Appendix C of~\cite{Ch91}, where the integration is from
                  $\theta=-\pi/2$ to $\theta=+\pi/2$. This interval corresponds to the full range of values of $\theta$ 
                  (modulo $2\pi$) for which $F_{cl}(q,\theta)$ exists when $q>0$. For $q<0$, $F_{cl}(q,\theta)$ exists 
                  when $\pi/2< \theta\,\mbox{mod}\,2\pi <3\pi/2$. However, the integrands of the integral representations 
                  in Eqs.~(C.62a,b) of~\cite{Ch91} are periodic with period $\pi$. (The definition of the variable 
                  $\theta$ in~\cite{Ch91} and in the present paper is such that $\theta=0$ when $p=0$ and $q>0$.) 
\bibitem{thetaPT} In the case of the P\"{o}schl-Teller potential, $F_{cl}(q,\theta)$ exists for $-\psi_q<\theta\,\mbox{mod}
                  \,2\pi<+\psi_q$ when $q>0$ and for $\pi-\psi_q<\theta\,\mbox{mod}\,2\pi<\pi+\psi_q$ when $q<0$, 
                  where $\psi_q\equiv\pi/2-(\pi/a)|q|$. These restrictions on $\theta$ emerge in the construction of 
                  $F_{cl}(q,\theta)$ as the Legendre transform of Hamilton's characteristic function for this potential.
\bibitem{FTtrick} Writing the gaussian $e^{-\alpha y^2}$ in terms of its Fourier transform, one can prove that 
                  \[\exp\left(-\frac{1}{4}\frac{\partial^2\ }{\partial y^2}\right)y^\varrho e^{-\alpha y^2}=
                    \frac{y^\varrho}{(1-\alpha)^{\varrho+1/2}}\exp\left[-\frac{\alpha y^2}{1-\alpha}\right]\]
                  for $\alpha>0$ and $\varrho=0,1$. The analytic continuation of this identity to arbitary $\alpha\ 
                  (=-z)$ is invoked in the steps leading to Eq.~(\ref{HOKw}).  
\bibitem{Fe98} F. M. Fern\'andez, Phys.\ Lett.\ A 237 (1998) 189.
\bibitem{PR82} I. Percival and D. Richards, {\sl Introduction to Dynamics}, (Cambridge University Press, Cambridge, 1982).
\bibitem{FPTalt} The expression in Eq.~(\ref{FPT}) is a little disconcerting in as much as it is not immediately apparent 
                 that $F_{cl}(q,\theta)$ is real-valued. However, it is possible to infer manifestly real expressions
                 if one considers $q$ of a definite sign and restricted ranges of $\theta$. For example, for $q>0$ and 
                 $|\theta|<\pi/2-\pi q/a$, Eq.~(\ref{FPT}) reduces to
                 $F_{cl}=J_s\left\{\theta-\mbox{Arcsin}[\sin\theta/\cos(\pi q/a)]\right\}$,
                 where Arcsin denotes the principal value of the inverse sine function.
\bibitem{ThetaPD} Calculation of this partial derivative starting from Eq.~(\ref{FPT}) is awkward. Instead, one can take
                  advantage of the fact that the expression in Eq.~(\ref{FPT}) is an even function of $q$ (for fixed 
                  $\theta$) and periodic in $\theta$ with period $\pi$ (for fixed $q$) to use the form of $F_{cl}(q,\theta)
                  $ given in~\cite{FPTalt}. Equation (\ref{Jqtheta}) for $J(q,\theta)$ can also be obtained without 
                  knowledge of $F_{cl}(q,\theta)$ by using the relation for $q$ in terms of angle-action variables, namely 
                  $q=(a/\pi)\mbox{Arcsin}\left([1-1/(1+J/J_s)^2]^{1/2}\cos\theta\right)$.
\bibitem{Overlap} If $\Theta$ is the angle between two wave functions defined using inner products, then $\cos\Theta$ is 
                  of the order of $1/\Lambda$. 
\bibitem{AS64} Ultraspherical or Gegenbauer polynomials are discussed in, for example, M. Abramowitz and I. A. Stegun 
              (eds.), {\sl Handbook of Mathematical Functions}, (National Bureau of Standards, Washington, 1964), ch.\ 22.
\bibitem{Mo02} J. Al-Modhayan, M.S. dissertation, Kuwait University, 2002, unpublished.
\bibitem{WKBnrm} Approximate normalization is achieved by employing the simple estimate given in Eq.~(2.78) in 
                 M. Brack and R. K. Bhaduri, {\sl Semiclassical Physics\/}, (Addison-Wesley, Reading, MA, 1997).
\bibitem{ExactHO} In making this demand, I am implicitly taking advantage of the fact that the semiclassical method
                  under consideration is exact for the simple harmonic oscillator.
\bibitem{MValue} The Morse oscillator generating function $F_{cl}(q,\theta)$ is single-valued for $q>0$ but double-valued 
                 for $q<0$. For $q>0$, $0\le\theta\,\mbox{mod}\,2\pi<\pi$ while, for $q<0$, $\psi_{cr}\le\theta\,
                 \mbox{mod}\,2\pi<\pi$, where, in terms of $\lambda_{cr}\equiv\sqrt{e^{-q/d}-1}$, $\psi_{cr}$ ($>\pi/2$) is
                 such that $\tan(\psi_{cr}/2)=(1+\lambda_{cr})/(1-\lambda_{cr})$. [As $q/d>-\ln 2$ for bound trajectories, 
                 reality of $\lambda_{cr}$ is guaranteed.] 
\bibitem{ConHO} This choice of branch implies that the term $\widetilde{\Lambda}e^{-q/d}$ in the argument of the global 
                exponential factor $\exp[-(\widetilde{\Lambda}-n-1/2)q/d-\widetilde{\Lambda}e^{-q/d}]$ in Eq.~(\ref{Mwave})
                is preceded by a minus sign. As a result, the quadratic small $q$ approximation to this exponential 
                factor, taking into account only those terms in the argument linear in $\widetilde{\Lambda}$ ($\gg 1$), 
                is proportional to the quadratic small $q$ approximation to the gaussian factor $\exp[-m\widetilde{\omega}
                q^2/(2\hbar)]=\exp[-\widetilde{\Lambda}q^2/(2d^2)]$ common to all wave functions of the related harmonic 
                oscillator problem.
\bibitem{Li86} R. G. Littlejohn, Phys.\ Rev.\ Lett.\ 56 (1986) 2000.
\bibitem{ZK96} D. Zor and K. G. Kay, Phys.\ Rev.\ Lett.\ 76 (1996) 1990.
\bibitem{MK98} M. Madhusoodanan and K. G. Kay, J. Chem.\ Phys.\ 109 (1998) 2644.
\end{thebibliography}
\end{document}